\documentclass{aa}
\usepackage{graphicx}
\usepackage{times}
\usepackage{astron}
\usepackage{psfig}
\newcommand{\sqdegr}{\raisebox{0.65ex}{\tiny\fbox{$ $}}\,$^{\circ}$}

\def\hide#1{}

\newcommand{\CA}{{\bf C}alar {\bf A}lto {\bf D}eep {\bf I}maging {\bf S}urvey}

\begin{document}

\thesaurus{04(10.19.3; 08.12.3)}

\title{CADIS deep star counts: Galactic structure and the stellar
luminosity function}
\author{S. Phleps\inst{1} \and K. Meisenheimer\inst{1} \and
B. Fuchs\inst{2} \and C.Wolf\inst{1}}
\offprints{S.Phleps (phleps@mpia-hd.mpg.de)}
\institute{
 Max-Planck-Institut f\"ur Astronomie, K\"onigstuhl 17,
   D-69117 Heidelberg, Germany
\and
 Astronomisches Rechen-Institut, M\"onchhofstr. 12-14, 
   D-69120 Heidelberg, Germany}

\titlerunning{CADIS deep star counts}
\date{Received 2. 6.  1999 / Accepted 3. 12. 1999}
\maketitle

\begin{abstract}
In this paper we present the first results of deep star counts carried
out within the \CA, CADIS \cite{meise}.
Although CADIS was designed as an
extragalactic survey, it also attempts to identify the stars in the
fields in order to avoid confusion with quasars and compact galaxies.\\
We have identified a sample of about 300 faint stars ($15.5\la R\la
23$), which are well suited to study the structure of the Galaxy. The stars
lie in two fields with central coordinates $\alpha_{2000}=16^{\rm h}
24^{\rm m} 32.^{\rm 
s}3$, $\delta_{2000}=55^{\circ} 44' 32''$ (Galactic coordinates: $l=85\degr$,
$b=45^\circ$) and $\alpha_{2000}=9^{\rm h} 13^{\rm m} 47.^{\rm
s}5$, $\delta_{2000}=46^{\circ} 14' 20''$ ($l=175^\circ$,
$b=45^\circ$) (hereafter 16h and 9h field, respectively. The 
stars have been separated from galaxies by a classification scheme
based on photometric spectra and morphological criteria. Distances
were derived by photometric parallaxes. We are able to find stars up to
distances of $\approx 25$\,kpc above the Galactic plane. The vertical
density distribution of the stars shows the contribution of the 
thin disk, the stellar halo and the ``thick disk'' of the
Galaxy. We give quantitative descriptions of the components in terms
of exponential disks and a de Vaucouleurs spheroid. For the disk stars
we derive the luminosity function. It is equal within the errors to the
local luminosity function and continues to rise out to at least $M_V$
= 13. Implications for the mass function are briefly discussed.

\end{abstract}
\section{Introduction}
The stellar structure of the Galaxy has been studied by many groups using a
variety of methods, such as metallicity-age determinations 
 or studies of the kinematics of nearby halo and disk
stars (for a review see Norris 1998). Other discussions can be found
in Fuchs \&
Jahrei{\ss} (1998), or Fuhrmann (1998). The method predominantly used to
study Galactic strucure is the method of {\it starcounts}. This
provides a measurement of the density distribution of the
stellar component of the Galaxy \cite{GilReid,Reid2,Reid3,Gould}.\\ 
In the so called {\it standard model} (Bahcall \& Soneira 1980) the
vertical structure of the disk follows an exponential law
($\rho\sim e^{-z/h_z}$) with scaleheight $h_z$. As Gilmore \& Reid
(1983) showed, the data can be fitted much better by a superposition
of two exponentials. Whether or not this deviation from the single
exponential is due to a distinct population of stars is not
clear, although this is suggested by the different kinematics and
lower metallicities \cite{Free}. We will call the deviation ''thick
disk'', regardless of its unknown physical origin.\\    
Although the existence of the ``thick disk'' component seems well
established now \cite{Norris}, the proper values for the scaleheights
and lengths are not 
exactly determined yet, since in most studies the distance of observable stars
has been limited to a few kpc above the Galactic plane.\\ 
Another topic of debate has been whether the stellar 
luminosity function (hereafter SLF) of main sequence stars in the disk
declines beyond 
$M_V=12^{{\rm mag}}$. Most photometric studies find a
down-turn of the 
slope of the SLF at $M_V=12^{{\rm mag}}$ \cite{Stobie,Kroupa3,Gould},
whereas local observations of stars 
in the solar neighbourhood indicate a flat continuation to fainter
magnitudes \cite{Wielen,Venice}. Therefore, it is worthwhile to derive
the SLF from the CADIS starcounts.\\
The paper is structured as follows:  the \CA  ~is outlined in Sect.
2. Sect. 3 describes the preparation of the stellar data, Sect. 4 
deals with the density distribution of the stars, Sect. 5 with the SLF
and its implications for the mass function. Sect. 6 gives a brief
summary and an outlook on future prospects.\\ 
\section{The Calar Alto Deep Imaging Survey}
The \CA~ combines an emission line survey carried out with an imaging
Fabry-Perot interferometer with a deep multicolour survey using three
broad-band optical to NIR filters and up to eighteen medium-band
filters when fully completed. The combination of different observing
strategies facilitates not only the detection of emission line objects but
also to derive photometric spectra of all objects in the fields
without performing time consuming slit spectroscopy.  Details of the
survey and its calibration will be given in Meisenheimer et
al. (in preparation).\\ 
All observations were performed on Calar Alto, Spain, in
the optical wavelength region with the focal reducers CAFOS (Calar
Alto Faint Object Spectrograph) at the 2.2 m telescope and MOSCA
(Multi Object Spectrograph for Calar Alto) at the 3.5 m telescope, and
with the Omega Prime camera for the NIR observations.\\ 
As a byproduct of
the survey we obtain a lot of multi-color data about faint stars in
the Galaxy.  Although for the object classification (see below)
exposures in two broadband filters and seven medium-band filters (in
the case of the 9\,h field eight medium-band filters) are used, the
present analysis of the stellar component of CADIS is based only on
exposures in three filters, $R_C$ (central wavelength/width
$\lambda_c / \Delta\lambda = 649\,{\rm nm}/170$\,nm), $B_C$ ($\lambda_c =
461\,{\rm nm}/100$\,nm) and $I_{815}$ ($\lambda_c = 815\,{\rm nm}/32$\,nm). Exposure times
converted to the 2.2\,m telescope are given in Table \ref{exptime}.\\
\begin{table}
\caption{Exposure times for the three broadband filters used for the
analysis of the stellar component in in the two CADIS fields. Numbers
shown in italics indicate, that the exposures were actually 
obtained at the 3.5 m telescope and scaled to the 2.2 m telescope.} \label{exptime}
\begin{tabular}{l|cc}
Filter & $t_{exp,16h}$ & $t_{exp,9h}$ \\
\noalign{\smallskip}
\hline
\noalign{\smallskip}
$B_C$ & 6200  & {\it 10000} \\ 
$R_C$ & 5300  & 2800\\
$I_{815}$ & 30700 & {\it 1750}\\
\noalign{\smallskip}
\hline
\end{tabular}
\end{table}
The nine CADIS fields measure $\approx 1/30~\sq\degr$ each and are
located at high Galactic 
latitude to avoid dust absorption and reddening. In all fields the total
flux on the IRAS 100\,$\mu$m maps is less than 2\,MJy/sr which
corresponds to $E_{B-V} < 0.07$, so we do not
have to apply any color corrections. A second selection criterium for
the fields was that there should be no star brighter than $\approx 
16^{mag}$ in the CADIS $R$ band. In fact the brightest star in the two
fields under consideration has an $R$ magnitude of $15.42^{mag}$.\\
\subsection{Object detection and classification} 
Objects are identified on each of the deep images (superposition of 5
to 15 individual exposures) using the Source
Extractor software  {\bf SE}xtractor \cite{Bertin}, and the resulting
lists merged into a master catalogue.\\
Photometry is done using the program {\it Evaluate}, which has been
developed by Meisenheimer \& R\"oser (1986). Variations in
seeing in between individual exposures are taken into account, in order to get accurate colors.
For photometric calibration we use a system of ''tertiary'' standard stars in
the CADIS fields, which are calibrated with secondary standard stars
\cite{oke,eso} in photometric nights.\\ 
From the locus of the stars in the 8-dimensional color space we
conclude that the relative calibration between each pair of wavebands
is better than 3 \% for all objects with $R=22$.\\ 
Since one of the major goals of the survey is the classification of every
object found in all CADIS fields ($\approx 80\, 000$ to $100\, 000$ in
total), a 
classification scheme was developed which is based on template spectral energy
distributions (see Wolf 1998). The observed colors of every object are
compared with a color library 
of known objects, whose colors are obtained from synthetic photometry
performed on our CADIS filterset. The input library for stellar spectra  
was the Gunn \& Stryker (1983) catalogue.
For each object the probability to belong to a certain object class
(stars -- quasars -- galaxies) is computed.\\
Objects classified as stars have stellar
colors with  a likelihood of more than 75\%, and images the profile of
which does not deviate
significantely from that of well defined stars.\\ 
Details about the performance and reliability of the classification
are given in Wolf et al. 1999, and Wolf (in preparation).\\
With the current filter set and exposure times the classification is
reliable down to a limit of $R \simeq 23^{mag}$. This was checked by
spectroscopic follow-up observations of 245 arbitrarily chosen objects (55 stars, 153
galaxies and 20 quasars) with $R<23^{mag}$. One galaxy has been
classified as a star by it's colors, two quasars as galaxies and two
galaxies as quasars. The star counts are not significantly affected by
the misclassifications down to $R=23^{mag}$.
Thus we restrict our present analysis to stars with $R\le 23^{mag}$.\\   
\section{Preparation of the stellar data}
\subsection{Photometric parallaxes}\label{data}
\begin{figure}[h]
\centerline{\psfig{figure=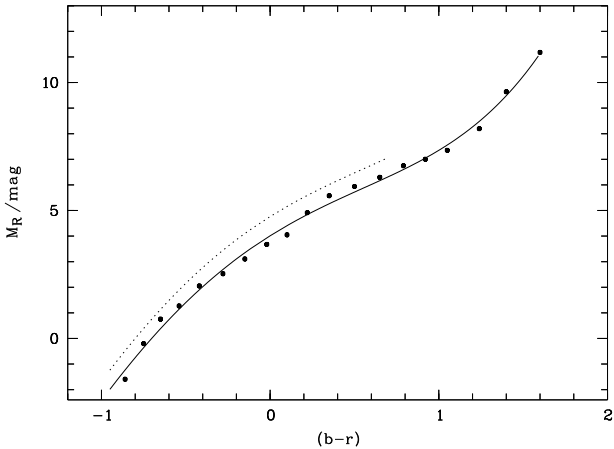,angle=270,clip=t,width=9.0cm}}
\caption[ ]{The mean main sequence from Lang (1992), converted to CADIS
color $b-r$ and $R_C$ magnitude; the solid line is the fourth order
polynomial (Eq. (\ref{polyfit})). For the blue ($b-r<0.7$) metal-poor stars 
the mean main sequence is shifted towards fainter magnitudes. \label{MSeq}}  
\end{figure}

\begin{figure}[h]
\centerline{\psfig{figure=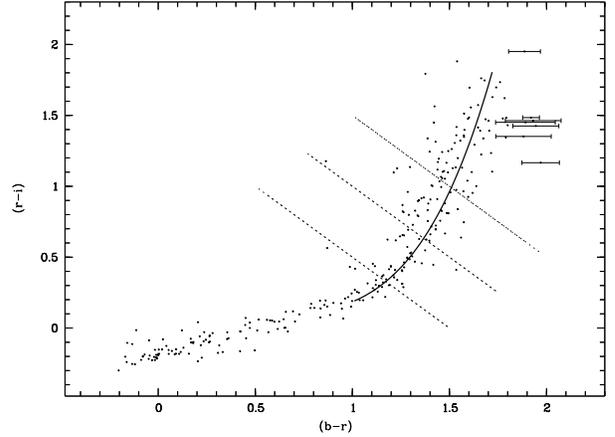,angle=270,clip=t,width=9.0cm}}
\caption[ ]{The $b-r$ colors are projected along the ''isophotes''
onto the mean main sequence track (solid line). For some of the very
red stars ($b-r>1.8$), which are subject to the largest corrections,
photometric errors are plotted.\label{metall}}   
\end{figure}

\begin{figure*}
\centerline{\psfig{figure=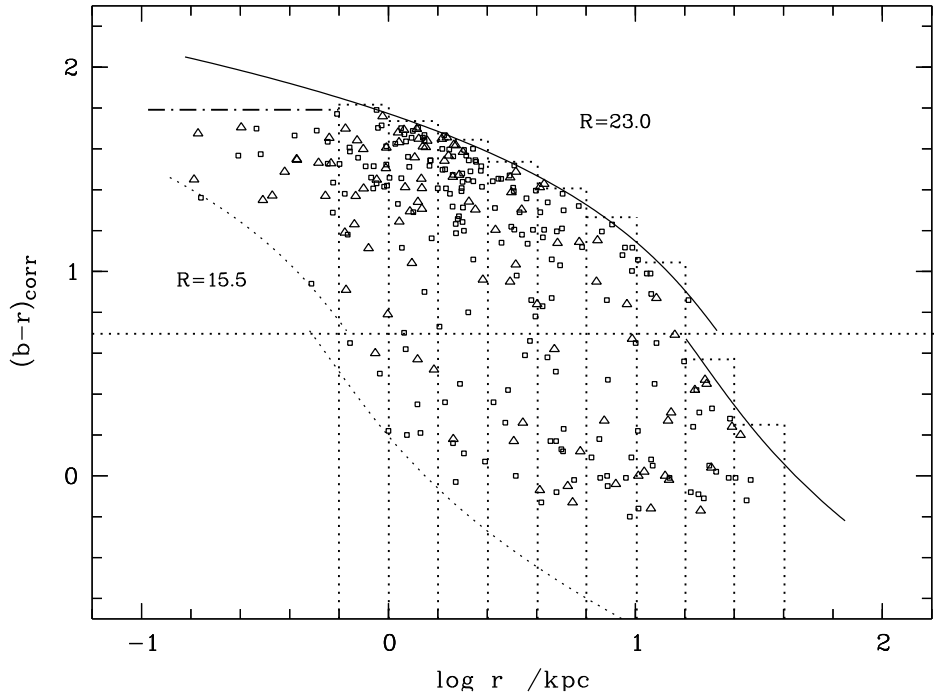,angle=270,clip=t,width=14cm}}
\caption[ ]{Spatial distribution of metallicity corrected
$(b-r)_{{\rm corr}}$.
The solid line represents
the distance dependent upper 
color limit at $R=23^{mag}$, the
dotted line is the lower limit due to the selection criterium of the
fields (no star brighter than $R\approx 15.5$). Since the metal-poor
halo stars are intrinsically fainter, the color limits are shifted
accordingly. The horizontal line
denotes the cut between halo and disk stars, the dashed--dotted line
indicates the cutoff at $(b-r)_{{\rm corr}}=1.8$ due to the metallicity
correction. The different symbols 
refer to the different fields: squares -- 16h field, triangles -- 9h
field. \label{wichtig}}
\end{figure*}

\begin{figure}[h]
\centerline{\psfig{figure=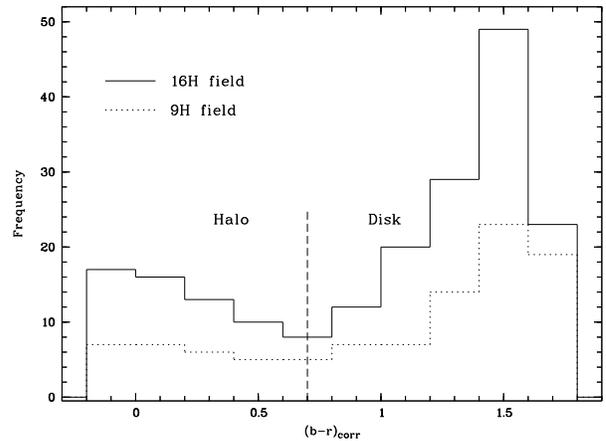,angle=270,clip=t,width=9.0cm}}
\caption[ ]{Distribution of $(b-r)_{{\rm corr}}$ in the two fields, this bimodal
distribution can be used to separate halo and disk stars by a color
cut. \label{codistr}} 
\end{figure}

To derive distances of the stars from the distance
modulus $m_R-M_R$, it is essential to know the absolute magnitudes
of the stars, $M_R$.\\
In principle absolute magnitudes of main sequence stars can be obtained
from a color-magnitude diagramm. This main sequence approximation is
valid for all stars in our sample since we can be sure that it is free
from contamination of any non-main sequence 
stars, as the faint magnitude intervall we observe ($16 \leq R \leq 
23$) does not allow the detection of a giant star.\\
We took a mean $M_{V_J}$
versus $(B-V)_J$ relation from Lang (1992). A complication arises,
because there is no $V$ filter included in our filter set. Thus we
have to convert $M_V$ and $(B-V)_J$ into our filter system, in order
to  derive the absolute magnitudes from mean main sequence fit in
$M_R$ versus $(b-r)$, where the CADIS color is defined by:
\begin{eqnarray}
(b-r)=2.5 \log \frac{F_R^\gamma}{F_B^\gamma}~.
\end{eqnarray}
Here $F^\gamma_k$ is the flux outside the atmosphere in units of Photons ${\rm
m}^{-2} ~{\rm s}^{-1} ~{\rm nm}^{-1}$ in the CADIS filter $k$.\\ 
In order to convert $(b-r)$ into the Johnson
$(B-V)_J$ we performed synthetic photometry on the Gunn-Stryker 
stars (Wolf, priv. comm.).\\
$b-r$ can be calibrated to the Johnson-Cousins system (the CADIS $R_C$ is
very close to the Cousins $R$) by using Vega as
a zero point:
\begin{eqnarray} 
(B-R)_C &=& (b-r) + 2.5 \log \frac{F^\gamma_{{\rm Vega}}(\lambda=440 {\rm
nm})}{F^\gamma_{{\rm Vega}}(\lambda=648 {\rm nm})}\nonumber\\
&=& (b-r)+0.725~.
\end{eqnarray}
The absolute magnitude $M_{R_C}$ is then given by
\begin{eqnarray} 
M_{R_C}&=&M_{V_J}-(V_J-R_C)~, {\rm with}\\
(V_J-R_C)&=&(B_C-R_C) - (B-V)_J~,\nonumber
\end{eqnarray}
where we assume $B_J = B_C$ (see Huang et al., in preparation).\\
With the above conversions, the main sequence ($M_R$ vs $(b-r)$) can
be approximated by a fourth 
order polynomial in the range $-1 \leq (b-r) \leq 1.8$: 
\begin{eqnarray}\label{polyfit}
M_R&=&c_0+c_1~(b-r)+c_2~(b-r)^2\nonumber\\
&+&c_3~(b-r)^3+c_4~(b-r)^4,
\end{eqnarray}
the parameters of which are:\\
\begin{center}
$c_0=4.01236$\hspace{0.6cm} $c_1=4.12575$\hspace{0.6cm}
$c_2=-1.89076$\hspace{0.6cm} $c_3=0.762053$\hspace{0.6cm}
$c_4=0.341384$~,\\
\vspace{0.3cm}    
\end{center} 
as shown in Fig. \ref{MSeq}.\\
One further complication arises due to the fact that the mean main sequence
relation is valid strictly only for stars with solar metallicities, whereas our
sample may contain stars spread over a wide range of different
metalicities. For the blue stars ($b-r<0.7$), which are almost all
halo stars (see Fig. \ref{codistr}), and therefore supposed to be metal-poor, we
use the same main sequence  
relation, but shifted towards fainter magnitudes by
$M_R=0.75^{mag}$ (see Fig. \ref{MSeq}). 
This value is the mean
deviation from the mean main sequence defined by the CNS   4 stars \cite{CNS4} of
a subsample of 10 halo stars, for which absolute R magnitudes 
were available (Jahrei{\ss}, priv. com.).\footnote{Note that this artificial
separation may well lead to wrong absolute 
magnitudes for individual stars (e.g. a disk star with $r<1$\,kpc but
$b-r<0.7$), but should be correct on average.}\\
The spread in a two
color diagram $(b-r)$ versus $(r-i)$ (that is the CADIS  color between
$R_C$ and $I_{815}$ analog to Eq. (1)) becomes significant at 
$(b-r) \approx 1.0$, see Fig. \ref{metall}. The maximal photometric error
for the very faint stars is $0.15^{mag}$.
Here metallicity effects will distort the relation 
between the measured $(b-r)$ colors and the spectral type (temperature) and
thus lead to wrong absolute magnitudes, so we have to correct for
metallicity in order to avoid errors in the photometric parallaxes.\\
The $R_C$
filter is strongly affected by metallicity effects like absorption
bands of TiO$_2$ and VO molecules in the stars' atmosphere, whereas
the $B_C$ and the medium-band filter $I_{815}$ (the 
wavelength of which was chosen in order to avoid absorption bands in
cool stars) are not. So in a
first approximation 
we can assume the ''isophotes'' of varying metallicity in a $(b-r)$
versus $(r-i)$ two color diagram to be straight
lines with a slope of 
$-1$, along of which we project the measured colors with $(b-r) \geq
1.0$ onto the mean main 
sequence track which in the interval $1.0 \leq (b-r) \leq 1.8$ is defined by
\begin{eqnarray}
(r-i)&=& 0.39~(b-r)_{{\rm corr}}^4  -0.36~(b-r)_{{\rm corr}}^3\nonumber\\
&+& 0.09~(b-r)+0.06~. 
\end{eqnarray}
This projection implies that stars with $(b-r)_{{\rm corr}}\ga
1.8$ cannot exist in
Fig. \ref{wichtig}, which shows the spatial distribution of metallicity
corrected $(b-r)_{{\rm corr}}$ colors (the limit is indicated by the
dashed--dotted line).  
Both the upper and lower magnitude limits lead
to selection effects which have to be taken into account. \\ 
As one can see in Fig. \ref{wichtig} there is an bimodality of the
observed color 
distribution. The accumulation of red stars in the upper left consists
mainly of disk 
stars, and is separated by a void from the blue stars,
which predominantly belong to the halo.   
Thus a crude disk-halo separation can be drawn by a color cut -- we take
stars with $(b-r)_{{\rm corr}}<0.7$ to be 
halo, stars with $(b-r)_{{\rm corr}}>0.7$ to be disk stars. In the following
we will make use of this color cut to derive the distribution of the
disk stars 
separately. In a second step, the distribution as a whole will be analysed. 
In Fig. \ref{wichtig} the cut is denoted by a dotted horizontal line.\\
In the two fields we
have analysed so far we find 95 halo and 178 disk stars, that is a factor
of two more disk stars than predicted by the standard model 
\cite{BSStandard}.  This 
surplus of disk stars was already noted by Reid \& Majewski
(1993). Fig. \ref{codistr} shows the distribution of the
$(b-r)_{{\rm corr}}$ in the two fields under consideration. 
\section{Density distribution of the stars}
\subsection{Completeness correction}
As expected, the detection limit of
absolute magnitudes (colors) is distance dependent (see Fig. \ref{wichtig}).\\
We use the two-dimensional distribution of stars in the $(b-r)_{{\rm corr}}$
vs $\log r$ diagram (Fig. \ref{wichtig}) to correct for this
incompleteness in the following way:
First, we divide 
the distance in logarithmic bins of 0.2 as indicated in
Fig. \ref{wichtig} and count the stars up to the 
the upper color limit $(b-r)^{lim}$ (this is the distance dependent
color (luminosity) limit, up to which stars can be detected. The
metal-poor halo stars are intrinsically fainter (see paragraph 3)),
thus the limits are shifted accordingly).\\    
The nearest bins ($-0.8\leq \log r \leq -0.2$) are
assumed to be complete. For the incomplete bins we multiply
iteratively with a factor given by the ratio of complete to incomplete
number counts in the previous bin, where the limit for the
uncorrected counts is defined by the bin currently under examination ($j$):
\begin{eqnarray}\label{Ncorrect}
N_j^{{\rm corr}}=N_j \prod_{i=1}^{j-1} (1+\frac{N_i^{''}}{N_i^{'}})~,
\end{eqnarray}
where $N_j$ is the number of stars in bin $j$, $N^{'}$ is the number
of stars in the previous bin ($j-1$), up to the 
limit given by the bin $j$, and $N^{''}$ is the number of stars from
that limit up to the limit given by bin $j-1$, see also appendix.\\
With the poissonian errors $\sigma_{N}=\sqrt{N}$, 
$\sigma_{N^{'}}=\sqrt{N^{'}}$, and $\sigma_{N^{''}}=\sqrt{N^{''}}$  the
error of the corrected number counts becomes:
\begin{eqnarray}
\sigma^2_{N_j^{{\rm corr}}}&=&\sigma^2_{N_j}\prod_{i=1}^{j-1}
\left(1+\frac{N_i^{''}}{N_i^{'}}\right)^2\nonumber\\
&+&\sum_{i=1}^{j-1}\left[\left(\frac{1}{N_i^{'}}\right)^2\sigma^2_{N_i^{''}}+\left(\frac{N_i^{''}}{N_i^{'}}\right)^2\sigma^2_{N_i^{'}}\right]\nonumber\\
&\cdot&\prod_{m=1 \atop m\not= j}^{j-1}\left(1+\frac{N_i^{''}}{N_i^{'}}\right)
\end{eqnarray}
For a detailed deduction see appendix.\\
The completeness correction is
done for each field separately.\\
With the corrected number counts the density in the logarithmic spaced
volume bins ($V_j=\frac{1}{3}\omega (r_{j+1}^3 - r_j^3)$) can than be
calculated according to   
\begin{eqnarray}
\rho_j=\frac{N^{{\rm corr}}_j}{V_j} =\frac{c \cdot N_j}{V_j}~,
\end{eqnarray}
For every logarithmic distance bin we use the mean height $z$ above
the Galactic plane $<z_j>=\sin b \cdot <r>$, where $<\log r> = \log
r_j+(\log r_{j+1}-\log r_i)/2=\log r_i+0.1 \Rightarrow <r>=1.259 r_i$.\\  
\subsection{Vertical density distribution}\label{VertDistr}
\begin{figure}[h]
\centerline{\psfig{figure=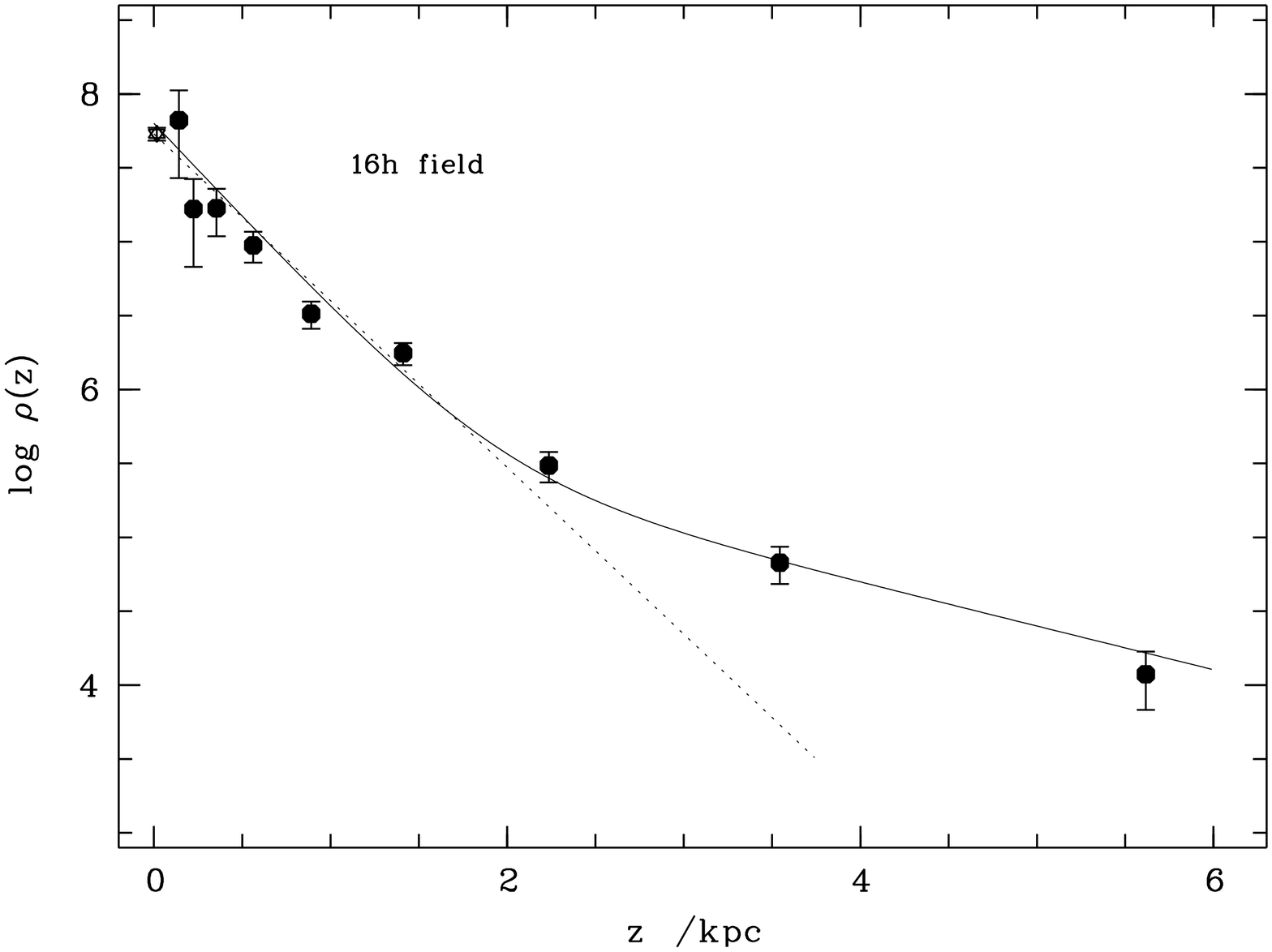,angle=270,clip=t,width=9.0cm}}
\centerline{\psfig{figure=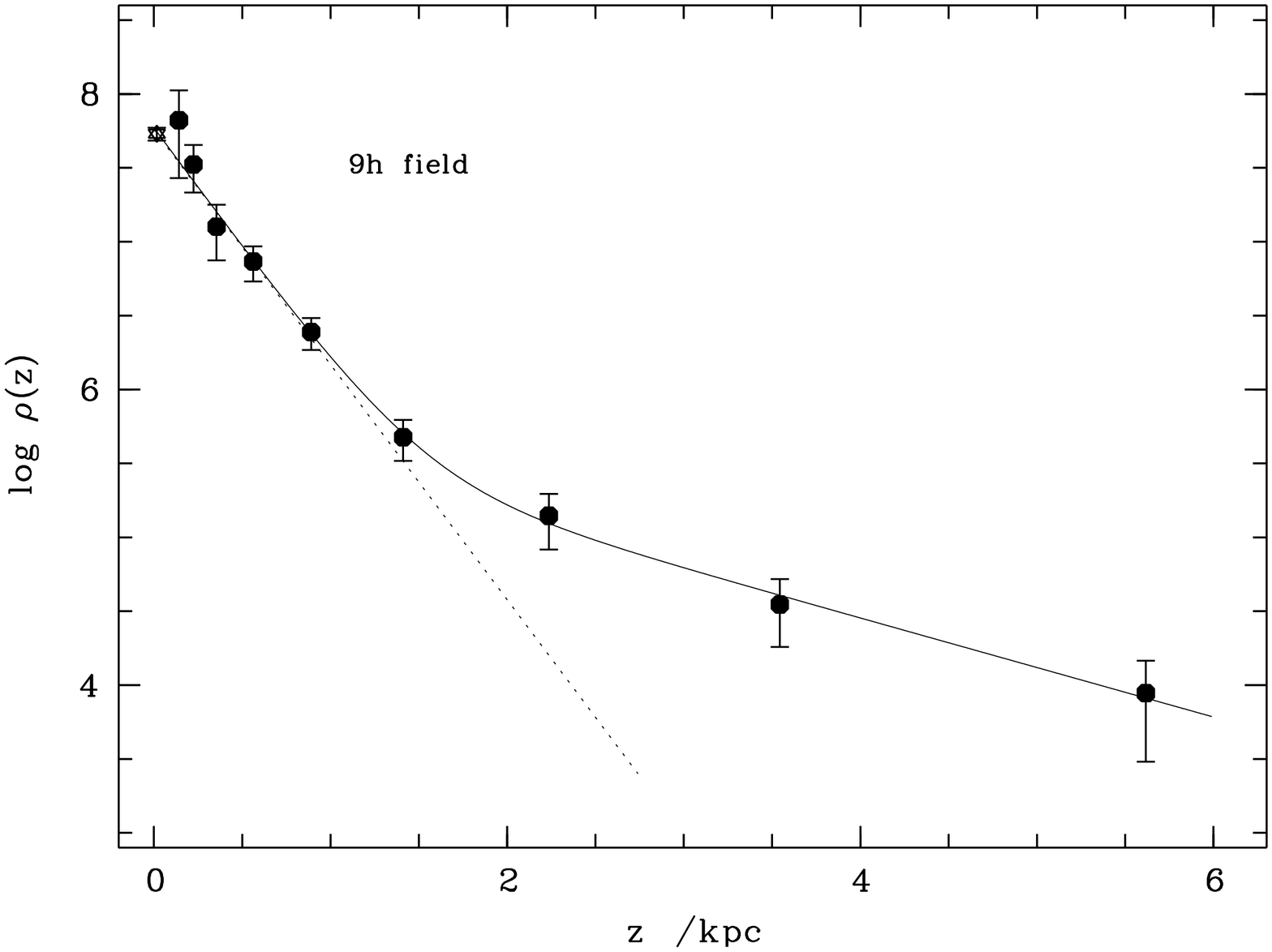,angle=270,clip=t,width=9.0cm}}
\caption[ ]{Vertical density distribution of the disk stars
($b-r>0.7$) in the two fields, the normalisation (star) is given by
data from the 
CNS4. The distribution shows clearly the contribution of the 
thick disk component -- the solid line is a fit with a superposition
of two exponentials, the dotted line is the single exponential fit for
the thin disk component only.  \label{Disk}}
\end{figure}

\begin{figure}[h]
\centerline{\psfig{figure=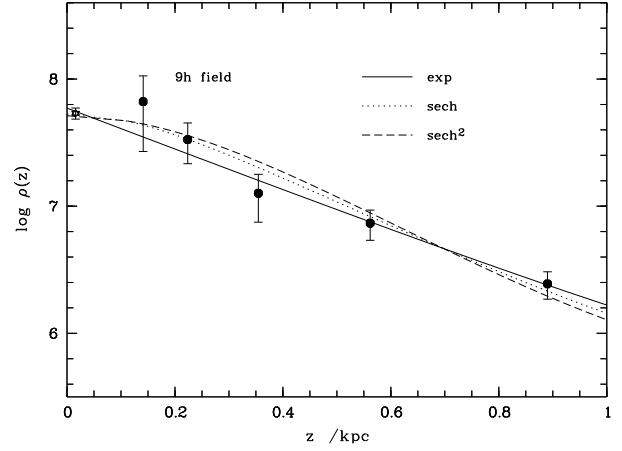,angle=270,clip=t,width=9.0cm}}
\caption[ ]{A secans hyperbolicus (dotted line) or a squared secans
hyperbolicus (dashed line) can be shown to fit the
thin disk equally well like an exponential (solid line). The
normalisation is given by data from the CNS4. \label{sech}}
\end{figure}

We first study the density distribution of the disk stars by taking
only stars in the corresponding color interval into
account. Although the color-cut at $(b-r)_{{\rm corr}}=0.7$ is a rather crude
separation between disk and halo, we gain a clearer insight into
the disk distribution since the contamination by halo stars is
suppressed considerably.\\    
As the nearest stars in our fields have still distances of about
200\,pc the normalization at $z=0$ has to be established by other
means: We take  stars from the CNS4 \cite{CNS4}, which 
are located in a sphere with radius 20\,pc around the sun. \\
The stars in our normalization sample 
are selected from the CNS4 by their absolute visual magnitudes,
according to the distribution of absolute magnitudes of the CADIS disk
stars ($6.5\leq M_v \leq 14.5$).\\
 Fig. \ref{Disk} shows the resulting density distribution of the disk stars in
the two CADIS fields. The solid line represents a fit with a superposition of
two exponentials, the dotted line is the fit for the thin disk
component (the first seven data points). Obviously a single
exponential is not a good description.\\
It was suggested  to fit the thin disk with a secans hyperbolicus --
the exponential is unphysical in that sense, 
that it is not 
continuously differentiable at $z=0$.  A squared secans
hyperbolicus (which represents a self-gravitating isothermal disk) can
be proved not to fit the data very well 
-- indicating that the stellar disk is in no  
way isothermal (the velocity dispersion depends on the spectral
type).\\
The fits for the three functions under consideration:
\begin{eqnarray}
\rho_{exp}(z) & = & n_1 \exp(-z/h_1)+n_2\exp(-z/h_2)\\
\rho_{sech}(z) & = & n_3 {\rm sech}(-z/z_0)+n_4\exp(-z/h_3)\\
\rho_{sech^2}(z) & = & n_5 {\rm sech}^2 (-z/\tilde{z_0})+n_6\exp(-z/h_4)~.
\end{eqnarray}
are shown in Fig. \ref{sech}, the corresponding
parameters are given in Table \ref{scales}.\\
 As the secans hyperbolicus is a sum of two exponentials
\begin{eqnarray}
{\rm sech}(z/z_0)=\frac{2}{\exp(z/z_0)+\exp(-z/z_0)}~,
\end{eqnarray}
$z_0$ is not really a scaleheight, but has to be compared to $h_1$
by multiplying it 
with arcsech$\frac{1}{{\rm e}}\approx 1.65745$: $h_1'= z_0\cdot
1.65745$
(cf. Tab. \ref{scales}).\\
\begin{table}[h]
\caption{Fit parameter for the three fit functions \label{scales}}
\begin{tabular}{lcc}
Parameter & 16h field  & 9h field\\
\noalign{\smallskip}
\hline
\noalign{\smallskip}
$n_1$ & 5.06$\cdot 10^7$ & 5.84 $\cdot 10^7$  \\ 
$n_2$ & 4.14 $\cdot 10^6$ & 6.23 $\cdot 10^5$  \\ 
$h_1$ & (280 $\pm$ 14)pc  & (267 $\pm$ 9)pc  \\ 
$h_2$ & (1267 $\pm$ 74)pc  & (1296 $\pm$ 116)pc  \\ 
\hline
$n_3$ & 4.54 $\cdot 10^7$ & 5.19 $\cdot 10^7$  \\ 
$n_4$ & 9.86 $\cdot 10^6$ & 1.02 $\cdot 10^6$  \\
$z_0$ & (172 $\pm$ 12)pc  & (216 $\pm$ 8)pc  \\ 
$h_1'$ & (286 $\pm$ 20)pc  & (358 $\pm$ 13)pc  \\ 
$h_3$ & (1478 $\pm$ 30)pc  & (1132 $\pm$ 84)pc  \\
\hline 
$n_5$ & 4.62 $\cdot 10^7$ & 5.4 $\cdot 10^7$  \\
$n_6$ & 2.78 $\cdot 10^6$ & 1.37 $\cdot 10^6$  \\
$\tilde{z_0}$ & (352 $\pm$ 26)pc  & (360 $\pm$ 13)pc  \\ 
$h_4$ & (1158 $\pm$ 112)pc  & (1047 $\pm$ 71)pc  \\ 
\noalign{\smallskip}
\hline
\end{tabular}
\end{table}
The errors of the scaleheights and corresponding parameters $z_0$ and
$\tilde{z_0}$ are
estimated by changing their values until $\chi^2$ increases
by 1. We find that at $z=0$ the contribution of the thick disk
component to the entire disk
is $2\pm 4 \%$.\\   
Our values for the scaleheights lie within the range given by different
authors in the literature. The proposed 
values for the exponential scaleheights of the thin disk range between
200 pc \cite{Kent} and 325 pc \cite{BSStandard}.
The parameters $\tilde{z_0}$ and $h_4$ we found in the 16\,h field are
equal within the errors to the values found by Gould et al (1997).\\
\subsection{Density distribution in the halo}
\begin{figure}[h]
\centerline{\psfig{figure=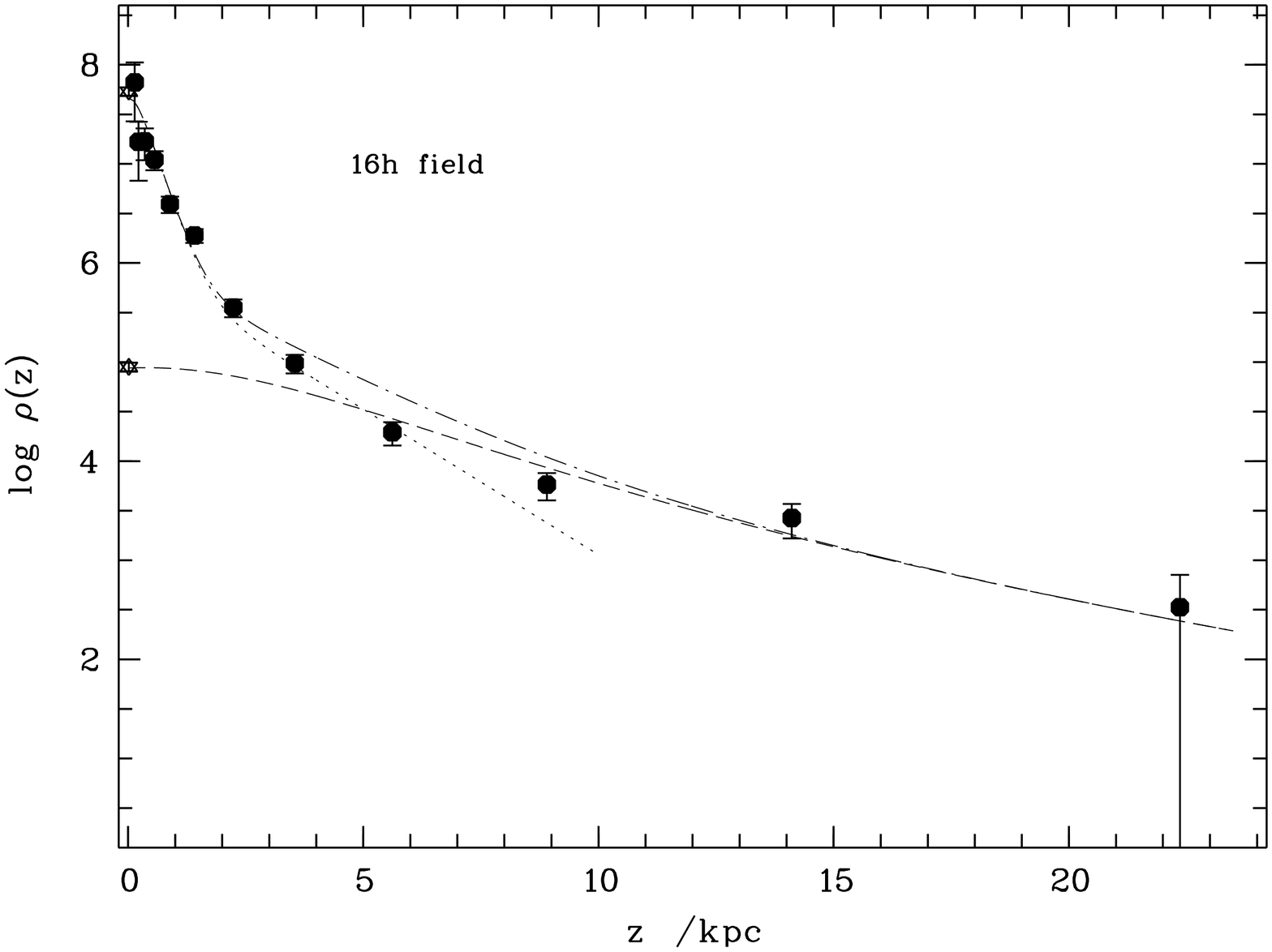,angle=270,clip=t,width=9.0cm}}
\centerline{\psfig{figure=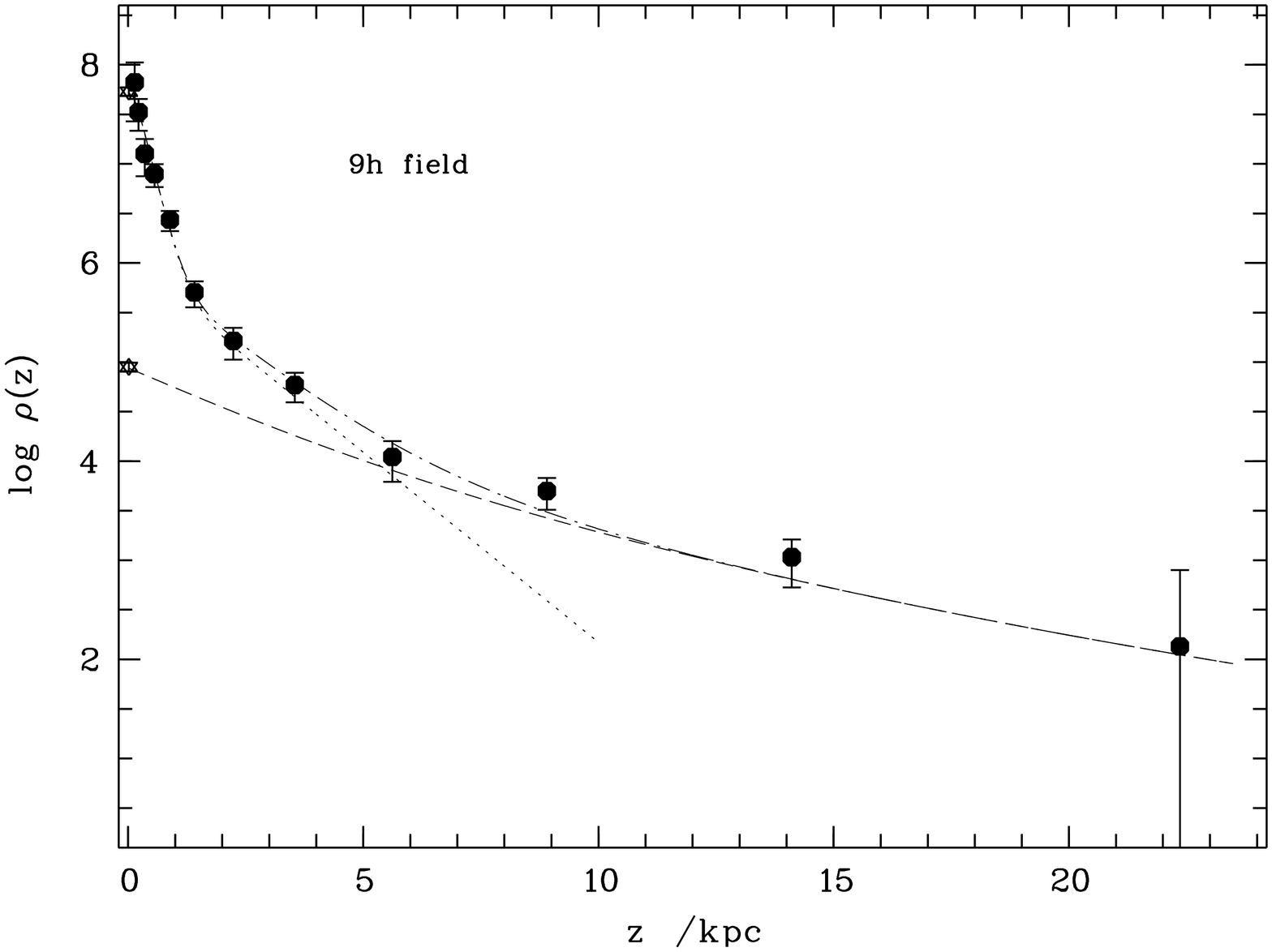,angle=270,clip=t,width=9.0cm}}
\caption[ ]{Density distribution perpendicular to the Galactic Plane,
stars are data taken from the CNS4; the dotted line is the
secans hyperbolicus plus exponential fit for the
disk, the dashed line is a deVaucouleurs law and the
dot-dashed the sum of both. \label{Dichte}}
\end{figure}

\begin{figure}[h]
\centerline{\psfig{figure=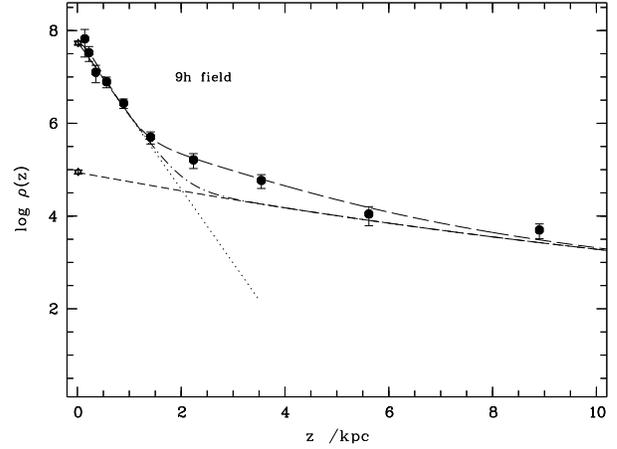,angle=270,clip=t,width=9.0cm}}
\caption[ ]{Distribution of all stars in the range
$0\leq<z\leq10$\,kpc, the fit for a single exponential disk (dotted line) the
deVaucouleurs law (short-dashed), the sum of both (dot-dashed), and
the sum of the thin plus thick 
disk fit and the deVaucouleurs law (long-dashed).\label{Sumtest}}
\end{figure}

\begin{figure}[h]
\centerline{\psfig{figure=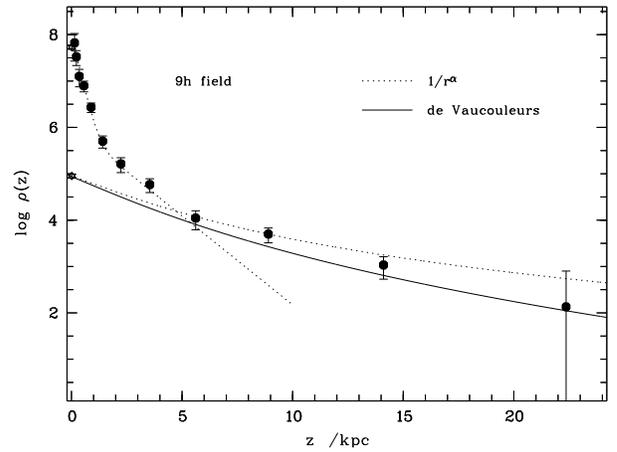,angle=270,clip=t,width=9.0cm}}
\caption[ ]{An inverse power law (broken line) can also be fitted to
the data -  in comparison with the a de Vaucouleurs law (solid
line), shown here for the CADIS 9h field. \label{Halotest}} 
\end{figure}

Fig. \ref{Dichte} shows the density distribution of {\it all} stars in the
two CADIS fields. The dotted line is the
secans hyperbolicus + exponential fit for the
thin and thick disk components, the dashed line is a deVaucouleurs
law. Although the space density corresponding to the $r^{1/4}$ law in
projection has no simple analytic form \cite{Young} it is possible to
form expansions about the origin and the point at infinity. The
latter expansion is valid for a large range of distances. An analytic
approximation is \cite{BSReview}:
\begin{eqnarray}\label{deVaucGneu}
&&\rho_H (z,b,l)=\rho_0\frac{\exp\left[-10.093\left(\frac{R}{R_\odot}\right)^{1/4}+10.093\right]}
{\left(\frac{R}{R_\odot}\right)^{7/8}}\nonumber\\
&\cdot&1.25\frac{\exp\left[-10.093\left(\frac{R}{r_\odot}\right)^{1/4}+10.093\right]}
{\left(\frac{R}{R_\odot}\right)^{6/8}}, R<0.03 R_\odot\nonumber\\
&\cdot&\left[1-0.08669/(R/R_\odot)^{1/4}\right], R\geq0.03 R_\odot\nonumber\\
\end{eqnarray}
where $R=(R_\odot^2+z^2+\frac{z^2}{\tan^2
b}-2 R_\odot \frac{z} 
{\tan b} \cos l)^{1/2}$, $b$ and $l$ are Galactic latidude and
longitude, the distance of 
the sun from the Galactic 
center $R_\odot=8$\,kpc,  $R=(x^2+z^2)^{1/2}$, $x=(R_\odot^2+d^2 
\cos^2 b-2R_\odot d \cos b \cos l)^{1/2}$, $z=d \sin b$.
The
dot-dashed is the sum of both halo and disk distribution
function.\\
The density of halo stars extrapolated to $z=0$ is compared with the data from
the CNS4 with an estimated error of $\approx 10$\%. The stars are discriminated
against disk stars 
by their metallicities and kinematics \cite{Fuchs}. They are further selected
according to the color cut of the CADIS halo stars ($(b-r\leq
0.7$).\\ 
Fig. \ref{Sumtest} shows the distribution of all stars in the range
$z=0$ to $z=10$\,kpc, the fit for a single exponential disk, the
deVaucouleurs law, the sum of both, and the sum of the thin plus thick
disk fit and the deVaucouleurs law. It is obvious that between
$\approx 1.5$\,kpc to 5.0 \,kpc the thick disk 
provides the predominant contribution to the overall distribution. Thus, the
sum of thin disk and deVaucouleurs halo does not fit the distribution.\\
If we assume the sun to be 8\,kpc away from the Galactic center and
the normalisation $\rho_0$ to be fixed by the data from the CNS4, all
parameters are completely determined. As can be seen in
Fig. \ref{Dichte}, the corresponding plots fit the distribution of  
the halo stars up to a distance of about 22\,kpc above the Galactic
plane.\\ 
The de Vaucouleurs law is an empirical description of the
density distributions of stars in ellipticals, bulges of spirals and
galactic halos \cite{deVauc} but it is equally justified to fit other distribution
functions to the data: a smoothed power law fits the halo stars equally well
\cite{Fuchs,Gould}, see e.g. Fig. \ref{Halotest}.
When applying the power-law fit, we add  a
further free parameter which describes the flattening of the halo:
\begin{eqnarray}
&&\rho_H(z,b,l)=\nonumber\\
&&\rho_\odot\!\left(\frac{r_c^2+r_\odot^2}{r_c^2+r_\odot^2+\frac{z^2}{tan^2
b}-2 r_\odot \frac{z}{\tan b} \cos l+\frac{z^2}{(c/a)^2}}\right)^{\alpha/2}\!,
\end{eqnarray}
with an arbitrarily chosen core radius of $r_c=1$\,kpc.\\
We find an axial ratio $c/a$ of $(0.63\pm 0.07)$, with an exponent
$\alpha=(2.99\pm 0.13)$.\\
\section{The stellar luminosity function}
\begin{figure}[h]
\centerline{\psfig{figure=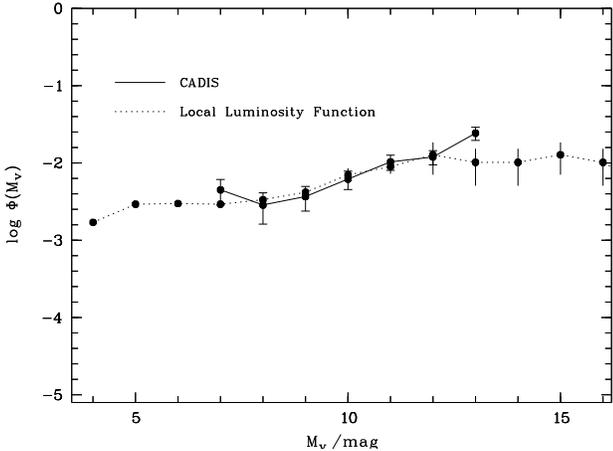,angle=270,clip=t,width=9.0cm}}
\caption[ ]{The mean SLF of the two CADIS fields (solid line) and the
local SLF \cite{Venice}. \label{LKF}}
\end{figure}

\begin{figure}[h]
\centerline{\psfig{figure=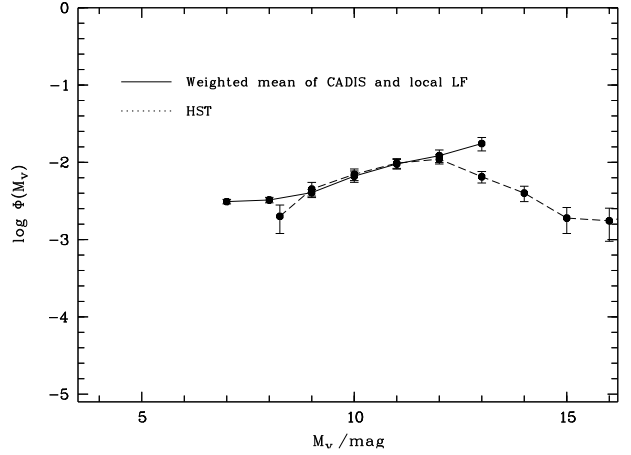,angle=270,clip=t,width=9.0cm}}
\caption[ ]{The weighted mean of CADIS and local SLF in comparison
with a photometric SLF which is based on HST observations
\cite{Gould}. \label{LKF_mean}} 
\end{figure}

The knowledge of the distribution function enables us to calculate the
SLF for the thin disk stars. We selected the stars with distances less than
1.5 kpc. Beyond this point the contribution by thick disk and halo stars becomes
dominant.\\ 
Although CADIS does not include a filter which is close to
the Johnson-V, we calculate absolute visual magnitudes from the $R$
magnitudes to make comparison with literature easier. In the relevant
magnitude intervall ($5 \leq M_V \leq 14$, i.e $-1 < (b-r) <
1.8$) there holds a linear relation (see Sect.
\ref{data}):
\begin{eqnarray}
M_{V_J}= 1.058 \cdot M_{R_C}~.
\end{eqnarray}
We calculate an effective 
volume for every luminosity bin by integrating the distribution
function along the line of sight, $r$, where the integration limits
are given by the minimum between 1.5\,kpc and the distance modulus
derived for upper and lower limiting apparent magnitude:
\begin{eqnarray}\label{vol}
V_{{\rm max}}^{{\rm eff}}= \omega \int\limits_{R_{{\rm min}}}^{R_{{\rm
max}}}
\nu(r,b) ~r^2 ~dr~,
\end{eqnarray}
where 
\begin{eqnarray*}
R_{{\rm min}}&=&10^{0.2(16^{{\rm mag}}-M_R)-2.0}~,\\
R_{{\rm max}}&=&{\rm min}(1.5~{\rm kpc}, 10^{0.2(23^{{\rm mag}}-M_{\rm
R})-2.0})~.
\end{eqnarray*}
The distribution function 
\begin{eqnarray}
\nu(r,b)= \exp(-r \sin b/ h_1)
\end{eqnarray}
is normalised to unity at $z=0$.\\  
The weighted mean of the SLFs of the two CADIS
fields is shown in Fig. \ref{LKF}, in comparison with the SLF of the
stars inside a distance of 20\,pc 
\cite{Venice}, which is based on HIPPARCOS parallaxes.
As can be seen from
Fig. \ref{LKF}, the CADIS SLF is equal within the errors to the local
SLF.  
Since the errors are dominated by Poissonian statistics, the error
bars are small at the faint end, and larger
at the bright end, complementary to the local SLF.\\ 
Thus the weighted mean of CADIS and local SLF which combines the large
number of bright stars in the local sample with our superior sampling
of faint stars (shown in Fig. \ref{LKF_mean}) can be regarded as 
the most  accurate determination of the SLF. The combined CADIS/local
SLF keeps rising with constant slope to its limit at $M_V=13$. This
is in pronounced contrast to previous determinations of the SLF based
on faint star counts \cite{Stobie,Kroupa3}. The discrepancy is
demonstrated in Fig. \ref{LKF_mean} where we compare the combined SLF
with the most recent photometric SLF based on HST observations
\cite{Gould}.\\
At this faint end both incompleteness and uncertainties of the exact
location of the main sequence may introduce systematic errors which
are hard to quantify.\\
\subsection{Implications for the mass function}
The stellar mass function (SMF) can in principle be inferred from the SLF
by converting the $M_V$ magnitudes of the disk stars ($r \leq
1.5$\,kpc) into masses; this requires a mass-luminosity relation, for
which we adopted the analytical fit taken from Henry et 
al. (1993, 1999). Their relation is valid for stellar masses from
$0.08\,M/M_\odot$ to $2.0\,M/M_\odot$. If we count the stars in equally
spaced mass intervals of $0.1$ and divide the number counts in the
$j_{th}$ interval by the maximum effective volume (see
Eq. (\ref{vol})), calculated for the corresponding luminosity,\\
the SMF can be represented by a power law:
\begin{eqnarray}
\Psi(M/M_\odot)\propto \left[\frac{M}{M_\odot}\right]^a~.
\end{eqnarray}
From the data of the disk stars in the two CADIS fields we derived
$a=-1.28\pm^{0.68}_{0.42}$ for [$0.2<M/M_\odot<1.1$]. Its slope
 is equal (within the errors) to the value proposed by Henry et
al. (1993) ($a=-1$).\\ 
\section{Summary and future prospects}
Although the \CA ~is designed as an extragalactic survey, 
we obtain a substantial set of multicolor-data about faint stars in the
Galaxy. With the current filter set and exposure times the
classification is reliable down to $M_R=23$.\\
From the $\approx 300$ stars we identified in two CADIS fields
covering a total area of 1/15 \sqdegr, we deduced the density
distribution of the stars up to a
distance of about 20\,kpc above the Galactic plane. The density
distribution shows 
unambiguously the contribution of a thick disk component of the Galaxy with a
scaleheight of $\approx 1.3$\,kpc.\\ 
There has been discussion whether this ''thick disk'' is introduced
artificially by the assumption that all stars are on the main sequence 
\cite{BS2}. Our present sample of very faint field stars does rule out
this ambiguity: if there were any
giant stars included in our data, 
they would have distances of about 250\,kpc, thus their contribution
to our sample can be neglected.\\
The density distribution in the halo, which essentially is corroborated by
comparing the local density of CNS4 stars with the density between
5.5 \,kpc and 15 \,kpc above the Galactic plane from CADIS, is
perfectly fit by a de Vaucouleurs' law.\\
Based on our present, limited data set of 72 stars beyond z=5.5 \,kpc,
it is equally well justified to fit an inverse power law with exponent
$\alpha=2.99\pm0.07$ to the data, 
with the axial ratio of the halo as an additional free parameter. 
The best fit value for the axial ratio is ($c/a = 0.63\pm 0.07$).\\
Based on this knowledge of the density distribution, we determined the
stellar luminosity 
function (SLF) for the disk stars. To this end, we confine our sample to stars with
distances less than 1.5\,kpc, for at this point the 
contribution of the thick disk and the halo
population becomes predominant. Our result is within the errors
indistinguishable from the
local SLF \cite{Venice}.  Thus we conclude that the weighted average
between the local SLF (containing predominantly stars at bright
magnitudes ($M_V<10$)) and the CADIS SLF (with superior statistics at
the faint end ($M_V>10$)) can be
regarded as the best estimate of the SLF of the disk stars. The SLF
continues to rise at least up to $M_V=13$, 
in contrast to most other photometric SLFs,
which show a down-turn at $M_V=12$.\\ 
We regard this as a hint that incompleteness corrections and the
closely connected issue of the ''true'' sampling volume of these faint
star counts might need some critical revision.\\
The stellar mass function (SMF) of the disk stars was derived by converting
the visual magnitudes into masses \cite{Henry1,Henry2}. For a
power-law SMF $\Psi \propto (M/M_\odot)^a$ we find a slope
$a=-1.28\pm^{0.68}_{0.42}$. This is (within the errors)
consistent with the 
slope proposed by Henry et al. (1993) ($a=-1$).\\
The main purpose of the present paper was to explore to what extent
the faint star counts in CADIS can be used for determining the stellar
density distribution and the stellar luminosity function of the
disk. However, one should keep in mind that our present analysis is
based only on about a quarter of  of the entire
\CA ~data. In the entire sample which will contain eight fields of 1/30
\sqdegr each, we expect to identify a total of $\ga 1200$ stars with
$R\leq 23$, of which 800 are supposed to be disk stars and 400 to be
halo stars.\\
In addition with the complete multicolor data of
the survey it should be possible to push the limit for a reliable star
classification to $R=23.5$ or beyond.\\
As soon as better template spectra of stars with sub-solar
metallicities become available, the currently rough correction for
metallicity effect can be replaced by a more accurate metallicity
dependent main sequence.\\
With this much more accurate determination of photometric distances
and an at least 4-fold increased statistics it should be possible to
adress the following issues concerning the density and luminosity
function of the stars in the Galaxy: 
\begin{itemize}
\item{the scalelength $h_r$ of the density distribution of stars in
the disk. This
can be done if one takes the longitude dependent radial decrease
of the density into account, e.g. the scaleheight measured in the CADIS 16h field
($l\approx 90\degr$) should be only very slightly affected by the
radial density gradient,
whereas the 9h field is suffering from a maximal radial
decrease in density, so one measures an effective scaleheight
$h_{{\rm eff}}$. Likely the star 
counts in the 18\,h and 23\,h field (at Galactic coordinates
$b=30\degr$, $l=95\degr$, and $b=-43\degr$, $l=90\degr$, respectively,
are only very slightly affected by the radial density 
gradient, so it should be possible to measure the ''true''
scaleheights $h_1$ and $h_2$ for thin and thick disk component with
high precision. From this, the counts in the 9h field and the
10\,h field ($b=53\degr$, $l=150\degr$), we 
will deduce the scalelength $r_i=[1/h_{{\rm eff}}-1/h]^{-1}$.}
\item{the axial ratio of a flattened halo and the exponent of the
inverse power law with higher precision.} 
\item{a bulge -- halo separation. By analysing the star counts in all
eight fields it will be possible to separate the contribution of the
bulge from the halo distribution.} 
\item{the slope of the SLF with higher precision. From the entire
data set with limiting magnitude $R\geq23.5$ we
will be able to deduce the SLF down to $M_V\geq 14$ with good statistics
and thus decide whether the features like the {\it 
Wielen dip} \cite{Wielen} at $M_V\approx 7$ and the apparent rise of
the SLF beyond
$M_V=12$ are real.}\\
\end{itemize}
\begin{acknowledgements}
The authors would like to thank H.Jahrei{\ss} who identified the
subset of stars from the Forth Catalogue of Nearby Stars ({\it CNS4})
and thus provided the local normalisation for the disk component.\\
We are
greatly indebted to the anonymous referee who pointed out several points
which had not received sufficient attention in the original
manuscript. This led to a substantial improvement of the paper.\\
We also
thank the Calar Alto staff for their help and support during many
observing runs at the observatory.
\end{acknowledgements}
\clearpage
\newpage
\appendix
\section{Completeness correction}
\begin{figure}[h]
\centerline{\psfig{figure=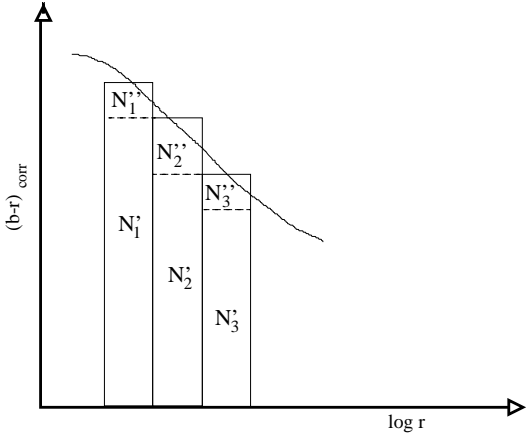,clip=t,width=7.0cm}}
\caption[ ]{Schematic sketch of Fig. \ref{wichtig}. \label{schematicCompCorr}}
\end{figure}

Since the first distance bin is assumed to be complete, we can
iteratively correct the number counts in the next bins.\\
$N_j$ is the number of stars in bin $j$, $N^{'}$ is the number
of stars in the previous bin ($j-1$), up to the 
limit given by the bin $j$, and $N^{''}$ is the number of stars from
that limit up to the limit given by bin $j-1$, as indicated in Fig.
\ref{schematicCompCorr}.\\
The division into $N^{'}$ and $N^{''}$ is done to avoid correlated
errors.\\
The first bin ($j=1$) is assumed to be complete ($N_1$). The corrected number
$N_2^{{\rm corr}}$ in the second bin is then
\begin{eqnarray}
N_2^{{\rm corr}}=N_2\cdot\left(\frac{N_1}{N_1^{'}}\right)\nonumber
=N_2\left(\frac{N_1^{'}+N_1^{''}}{N_1^{'}}\right)
=N_2\left(1+\frac{N_1^{''}}{N_1^{'}}\right)~,
\end{eqnarray}
and for the third bin
\begin{eqnarray}
N_3^{{\rm corr}}&=&N_3\left(\frac{N_2^{{\rm corr}}}{N_2^{'}}\right)=N_3\left(\frac{N_1}{N_1^{'}}\right)\left(\frac{N_2}{N_2^{'}}\right)\nonumber\\
&=&N_3\left(\frac{N_1^{'}+N_1^{''}}{N_1^{'}}\right)\left(\frac{N_2^{'}+N_2^{''}}{N_2^{'}}\right)\nonumber\\
&=&N_3\left(1+\frac{N_1^{''}}{N_1^{'}}\right)\left(1+\frac{N_2^{''}}{N_2^{'}}\right)~.\nonumber
\end{eqnarray}
Thus this can generally be written as
\begin{eqnarray}
N_j^{{\rm corr}}=N_j\prod_{i=1}^{j-1}\left(1+\frac{N_i^{''}}{N_i^{'}}\right)
\end{eqnarray}
The errors of the number counts $N_j$, $N_{j-1}^{'}$ and
$N_{j-1}^{''}$ are independent of each other and thus 
Poissonian: $\sigma_{N_j}=\sqrt{N_j}$,
$\sigma_{N_{j-1}^{'}}=\sqrt{N_{j-1}^{'}}$,
$\sigma_{N_{j-1}^{''}}=\sqrt{N_{j-1}^{''}}$. Following Gaussian error
propagation the error for the corrected number of stars in the second
bin is
\begin{eqnarray}
\sigma^2_{N_2^{{\rm corr}}}&=&\sigma^2_{N_2}+\left(\frac{N_1^{''}}{N_1^{'}}\right)^2\sigma^2_{N_2}\nonumber\\
&+&\left(\frac{N_2}{N_1^{'}}\right)^2\sigma^2_{N_1^{''''}}+\left(\frac{N_2
N_1^{''}}{N_1^{''2}}\right)^2\sigma^2_{N_1^{'}}~,\nonumber
\end{eqnarray}
and for the third bin 
\begin{eqnarray*}
\sigma^2_{N_3^{{\rm corr}}}&=&\sigma^2_{N_3}\left(1+\frac{N_2^{''}}{N_2^{'}}\right)^2
\left(1+\frac{N_1^{''}}{N_2^{'}}\right)^2+\\
&+&N_3^2\left(1+\frac{N_1^{''}}{N_1^{'}}\right)\left[\left(\frac{1}{N_2^{'}}\right)^2\sigma_{N_2^{''}}+\left(\frac{N_2^{''}}{N_2^{'}}\right)^2\sigma^2_{N_2^{''}}\right]\\
&+&N_3^2\left(1+\frac{N_2^{''}}{N_2^{'}}\right)\left[\left(\frac{1}{N_1^{'}}\right)^2\sigma_{N_1^{''}}+\left(\frac{N_1^{''}}{N_1^{'}}\right)^2\sigma^2_{N_1^{''}}\right]
\end{eqnarray*}
Thus the error of the completeness corrected counts in the $j^{th}$ bin is
\begin{eqnarray}
\sigma^2_{N_j^{{\rm corr}}}&=&\sigma^2_{N_j}\prod_{i=1}^{j-1}\left(1+\frac{N_i^{''}}{N_i^{'}}\right)^2\nonumber\\
&+&\sum_{i=1}^{j-1}\left[\left(\frac{1}{N_i^{'}}\right)^2\sigma^2_{N_i^{''}}+\left(\frac{N_i^{''}}{N_i^{'}}\right)^2\sigma^2_{N_i^{'}}\right]\nonumber\\
&\cdot&\prod_{m=1 \atop m\not= j}^{j-1}\left(1+\frac{N_i^{''}}{N_i^{'}}\right)~.
\end{eqnarray}

\end{document}